# Copper Vapor Catalyzing Role in the Growth of Graphene

Yuan Chang[a§], Shiji Li[a§], Tianyu He[a], Hongsheng Liu[a], Junfeng Gao[a,b]*

[a] *State Key Laboratory of Structural Analysis for Industrial Equipment, Dalian University of Technology, Dalian, 116024, China*

[b] *Suzhou Laboratory, Suzhou, 215123, China*

*Email: gaojf@dlut.edu.cn

[§] Y. C. and S. L. contributed equally to this work

**Abstract:** Cu is most used substrate to grow monolayer graphene under a temperature near melting point. In this study, we elaborated a remarkable amount of Cu clusters were continuously evaporated during the graphene growth, resulting into the vapor pressure comparable with the $CH_4$. Importantly, the decomposition barrier of $CH_4$ on Cu clusters is similar or even lower than on Cu surface. $CuCH_4$ serves as the primary active cluster in complex intermediates, exhibiting a growth-promoting effect. Particularly after the first graphene layer coverage, it may emerge as a dominant catalytic factor for multilayer growth by supplying critical growth species. Through controlled seed incorporation, this mechanism is expected to enable large-area controllable growth of bilayer and multilayer graphene structures.

Graphene is a standout material, widely used in various applications including seawater desalination, wearable devices, heat dissipation films, transparent touchscreens, ultra-sensitive sensors, biomedical diagnostics and therapy, and nano-drug delivery. Chemical vapor deposition (CVD) enables the bottom-up growth of graphene with large-area and precise controllability[1-3], which has emerged as the preferred method for scaled-up production. Graphene production via CVD relies on carbon-containing precursor like methane ($CH_4$) and catalytic substrate like Copper (Cu) due to its ability to facilitate dehydrogenation of hydrocarbon. The weak interaction and low carbon solubility of Cu enable direct single-layer graphene formation on its surface and exhibit self-limiting effect at low pressure [4].

Notably, Cu substrate can be evaporated and form Cu clusters vapor under the growth temperature near Cu melting points. Small clusters are usually very active and have excellent catalysis capacity in chemistry [5, 6]. We can expect that the Cu clusters in vapor are also active for C precursors ($CH_4$) in the CVD process. Using Cu foam [7] and Cu-containing carbon sources [8] ensures a continuous supply of Cu vapor, promoting ultra-clean single-layer graphene growth. Pre-depositing Cu allows graphene formation on inert $SiO_2$ substrate [9]. These initial efforts confirm the existence of non-negligible Cu clusters vapor and their important role in the CVD growth of graphene.

However, the role of Cu clusters in vapor is still quite unclear. What are the real structure and population of Cu clusters? How they interact with precursors? Can Cu clusters in vapor help the dehydrogenation of $CH_4$ and promote the graphene? What is the kinetics of $CH_4$ catalyzing by Cu clusters in vapor and their impact on gas-phase equilibrium? So many puzzles were poorly understood, and accurate comprehensive atomic simulations is urgent to unveil the underlying mechanisms.

In this letter, we systematically explored the conditions, structure, stability, kinetics, and role of Cu vapor in graphene growth during $CH_4$ decomposition using first-principles calculations. We first examined the adsorption-desorption behavior of Cu vapor by comparing chemical potentials and adsorption energies. The derived pressure–temperature (P–T) conditions align with most previous experiments, confirming the widespread presence of Cu vapor in the CVD environment for graphene growth. $Cu_{1\sim4}$ clusters, which account for 99% of all Cu clusters, exhibit catalytic activity comparable to, or even surpassing, Cu substrate surfaces in the initial $CH_4$ dehydrogenation. We identified the most stable $CuCH_4$ cluster for complex gas-phase reactions. Our analysis suggests that surface-adsorbed $CuCH_4$ clusters tend to diffuse towards the edge of pre-provided graphene nucleus on the top layer. This behavior may represent a growth mechanism that occurs once the Cu substrate is fully covered by graphene. Therefore, the generation and behavior of Cu vapor are critical for controlling the CVD process.

The adsorption–desorption behavior can be described by comparing the chemical potential ($\mu$) of ideal gas with its adsorption energy ($E_{ads}$). When $E_{ads}$ is less than $\mu$, net adsorption of atoms will continue, whereas desorption from substrate to vapor will occur when $\mu$ is less than $E_{ads}$. The $\mu$ for the ideal gas is defined as the following equations [10]:

$$\mu = -k_B T \ln(g k_B T/p \times \zeta_{trans}\zeta_{rot}\zeta_{vibr}) \tag{1}$$

$$\zeta_{trans} = (2\pi m k_B T/h^2)^{3/2} \tag{2}$$

$$\zeta_{rot} = (1/\pi\sigma)\{8\pi^3(I_A I_B \cdots)^{1/n} k_B T/h^2\}^{n/2} \tag{3}$$

$$\zeta_{vibr} = \prod i 3N - 3 - n\{1 - e^{-h\nu_i/k_B T}\}^{-1} \tag{4}$$

where $\zeta_{trans}$, $\zeta_{rot}$ and $\zeta_{vibr}$ are the partition function for the translational motion, the rotational motion and the vibrational motion, respectively. It is feasible to consider the Cu vapor as consisting of individual Cu atoms. This approach allows the complex vibrational entropy and rotational entropy to be disregarded. Subsequently, $\mu$ of Cu vapor, can be calculated:

$$\mu = -k_B T \ln(k_B T/p \times g(2\pi m k_B T/h^2)^{3/2}) \tag{5}$$

where $k_B = 1.381 \times 10^{-23}$ $J/K$ is the Boltzmann constant. $T$ is gas temperature, $p$ is beam equivalent pressure of particle, $g = 2$ is the degeneracy of electronic energy levels for Cu, $m = 1.063 \times 10^{-25}$ $kg$ is the mass of a particle, $h = 6.626 \times 10^{-34}$ $eV \cdot s$ is the Planck constant. It should be noted that actual vapor may contain Cu clusters, the total partial pressure of Cu vapor is higher than estimation.

Under equilibrium conditions, $E_{ads}$ of Cu vapor is defined as the following equation:

$$E_{ads} = E_{sub+*Cu} - E_{sub} - E_{isolate\ Cu} \tag{6}$$

where $E_{sub+*Cu}$ is the total energy of system, $E_{sub}$ and $E_{isolate\ Cu}$ denote the energy of Cu substrate and isolate Cu atom in vacuum, respectively. The actual copper substrate may consist of multiple crystal facets. In this study, we tested the $E_{ads}$ of Cu on different crystallographic planes. As shown in Fig. 1(a), our calculations indicate $E_{ads}$ of -3.382 eV for Cu(100), -3.746 eV for Cu(110) and -3.129 eV for Cu(111) surface. The adsorption sites had been carefully tested.

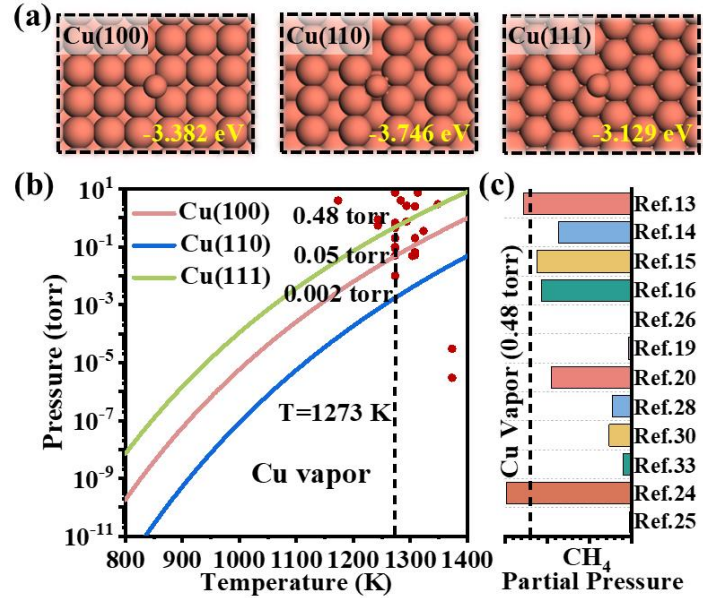

**Fig. 1.** (**a**) Adsorption energy of single Cu on common Cu surfaces. (**b**) The adsorption–desorption transition curve of Cu vapor. Red dots represent experimental P–T conditions [11-37]. (**c**) Comparison of Cu vapor and experimental CH$_4$ partial pressure at 1273K.

The adsorption–desorption transition curve of Cu vapor were given in Fig. 1(b), with the equilibrium partial ranking from highest to lowest as Cu(111) > Cu(100) > Cu(110). This can be easily explained by the adsorption strength of Cu atoms on different surfaces, among which Cu(111) has the weakest adsorption on Cu atoms. The statistical analysis of existing low-pressure experimental data [11-37] (Red dots in Fig. 1(b)) indicates that most P–T conditions satisfy the desorption criteria proposed in this study. In the experimental investigation of Cu contamination, it is evident to the naked eye that the Cu deposition on the inner walls of quartz tube after CVD growth is remarkably far broader than the area of the substrate [38, 39]. These evidences confirm the prevalence of Cu vapor in graphene synthesis (see Sec. 1 of the Supplemental Material for more details). As shown in Fig. 1(c), the Cu vapor can reach 0.48 torr at 1273 K, which is comparable to the partial pressure of CH$_4$ under the same temperature, emphasizing its significance as a non-negligible factor.

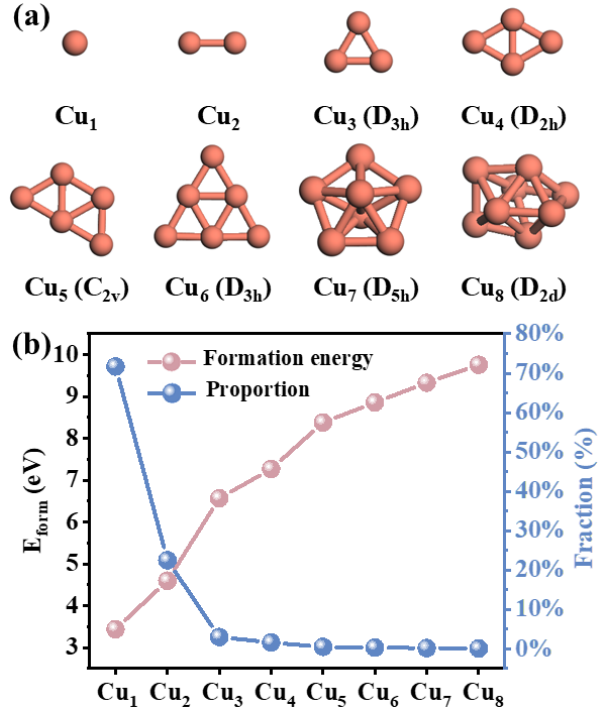

**Fig. 2.** (**a**) The ground state structures and (**b**) corresponding formation energy as well as proportion of Cu$_{1\sim8}$ clusters.

In the preceding discussion, Cu vapor was considered as individual atoms; however, in real growth conditions during graphene synthesis via CVD, collisions and aggregation of Cu atoms can occur. Therefore, determining the size of Cu clusters is essential before further research. The relative population of Cu clusters with different size can be estimated using the formula:

$$N \propto e^{-\Delta E/k_B T} \tag{7}$$

where $\Delta E$ is the relative energy difference of Cu cluster formation. To facilitate this analysis, the formation energy of Cu clusters was defined as:

$$E_{form} = E_{cluster} - N_{Cu}E_{Cu} \tag{8}$$

where $E_{cluster}$ represents the total energy of Cu cluster, $E_{Cu}$ denotes the energy per Cu atom in its face-centered cubic bulk phase and $N_{Cu}$ is the number of Cu atoms within the Cu cluster. As shown in Fig. 2, we had determined the ground state structures of Cu$_{1\sim8}$ clusters. The proportion of Cu$_{1\sim4}$ clusters has weighted 99%, while that of Cu$_{5\sim8}$ clusters has markedly reduced to less than 1% of the total content. This finding indicates a low probability of individual Cu atoms colliding and agglomerating into larger clusters in the

Cu vapor. Consequently, for the Cu cluster-catalyzed $CH_4$ dehydrogenation process being considered, the focus on $Cu_1$ to $Cu_4$ clusters is deemed adequate.

Due to the structural stability of $CH_4$ molecules, the initial step of dehydrogenation faces a challenging energy barrier [40]. Nevertheless, a series of radical chain reactions will be initiated facilitated by C–H bond activation after this step, resulting in the significant production of carbon radical groups. Therefore, it is imperative to concentrate specifically on the process where the first H atom is desorbed from the C atom. For Cu cluster catalysis as depicted in Fig. 3, the most abundant gas-phase $Cu_1$ and $Cu_2$ clusters in the growth chamber demonstrate improved barriers of 1.50 eV and 1.57 eV for the initial dehydrogenation of $CH_4$. Surprisingly, $Cu_3$ and $Cu_4$ clusters, with lower content, exhibit even better catalytic performance for the desorption of the first H atom in $CH_4$, with their barriers significantly reduced to 0.92 eV and 0.77 eV, respectively. We also considered CuH cluster with very stable electronic structure, which exhibits 1.73 eV barrier for the first step dehydrogenation of $CH_4$, indicating that hydrogenated Cu clusters are still catalytic (Fig. S6). For comparison, Cu(111) surface exhibits barrier of 1.69 eV, indicating that small Cu clusters exhibit more superior catalytic activity for the initial dehydrogenation of $CH_4$. We investigated the energy barrier for the subsequent dehydrogenation on the most abundant $Cu_1$ cluster as well. More details correspond to Cu cluster and surface catalysis were given in Sec. 2 of the Supplemental Material.

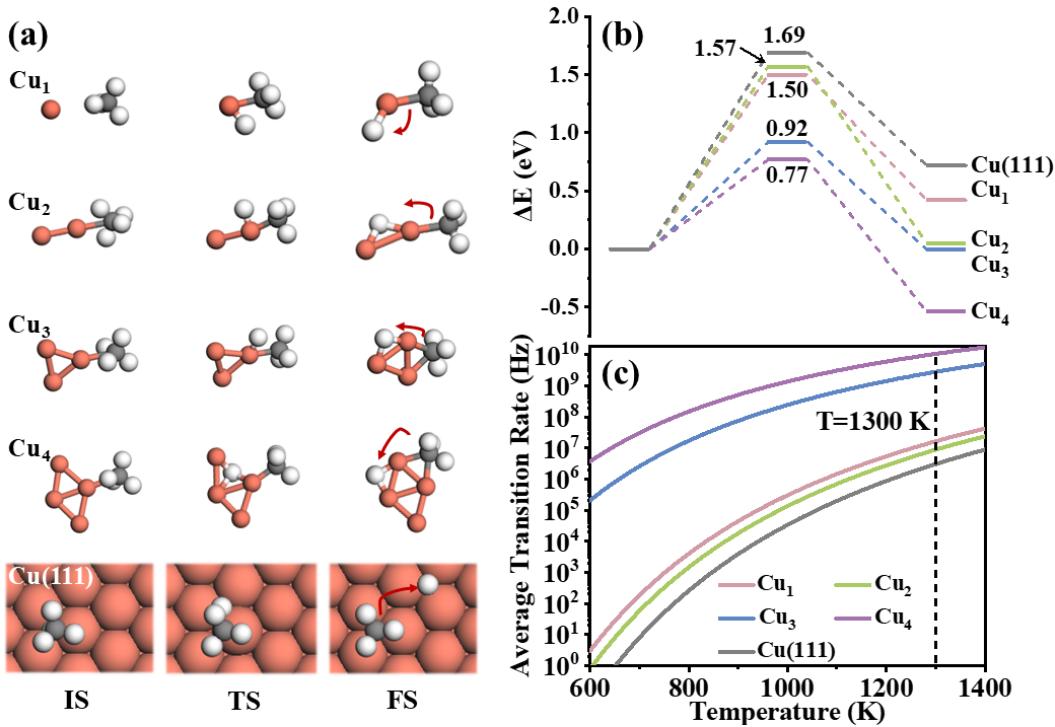

**Fig. 3.** (**a**) Atomic structures of initial state (IS), transition state (TS), and final state (FS). (**b**) Dehydrogenation barriers and (**c**) Average transition rates of Cu vapor.

The average transition rate for the initial desorption of H atoms from CH$_4$ can be approximated using the Arrhenius equation:

$$k = A \times e^{-E/k_B T} \quad (9)$$

where the prefactor $A$ is approximately equal to $10^{13}$ Hz, $E$ denotes the energy barrier. Fig. 3(c) illustrates the average transition rates of desorbed hydrogen atoms in Cu$_{1\sim4}$ clusters across temperatures ranging from 600 K to 1400 K. At the growth temperature of 1300 K, the atomic vibrational frequency associated with hydrogen atom desorption and migration processes on Cu$_1$ and Cu$_2$ clusters reaches $10^7$ Hz per second. On Cu$_3$ and Cu$_4$ clusters, this frequency can even exceed $10^9$ Hz. Such heightened vibrational frequencies strongly suggest that dehydrogenation processes are highly favorable with the assistance of Cu clusters. For comparison, the vibration frequency of Cu(111) surface is only $10^6$ Hz, significantly lower than that observed with Cu cluster catalysis. This once again highlights the significant role of isolated gas-phase Cu clusters in the CVD process of graphene growth. Despite the lower abundance of Cu vapor, their impact on gas-phase equilibrium and product control cannot be overlooked.

The assistance role of Cu clusters in the initial dehydrogenation of CH₄ had been elucidated through kinetic analysis. Under growth conditions, there are abundant CH₄ and H₂ in the growth chamber. Subsequently, thermodynamic discussion on the gas-phase equilibrium relationships during graphene growth will be focused. Here, the chemical potential $\mu(T,P)$ and $\chi = P_{CH_4}/P_{H_2}$ were introduced to represent the change in system free energy and CH₄–H₂ gas intake ratio, respectively. According to the NIST-JANAF thermodynamic tables [41] and the previous work [40], $\mu_H$ and $\mu_C$ can be derived as following equations:

$$\mu_H(T,P,\chi) = \mu_H(T,P_0) + \frac{1}{2}k_B T \ln \frac{P}{P_0} \frac{1}{1+\chi} \quad (10)$$

$$\mu_C(T,P,\chi) = \Delta g_{CH_4}(T,P_0) - 9.234 + k_B T \ln \frac{P}{P_0} \frac{\chi}{1+\chi} \quad (11)$$

where reference pressure $P_0 = 1\ bar$. Continuous temperature model was established by fitting the entropy and enthalpy from 800 K to 1400 K of H₂ and CH₄. Based on equations (5), (10), and (11), the symbolic expressions of $\mu_{Cu}(T,P)$, $\mu_H(T,P,\chi)$, and $\mu_C(T,P,\chi)$ regarding temperature $T$, pressure $P$ and gas intake ratio $\chi$ were obtained. Once specific growth conditions were measured from experiment, the values of $\mu_{Cu}$, $\mu_H$, and $\mu_C$ can be determined.

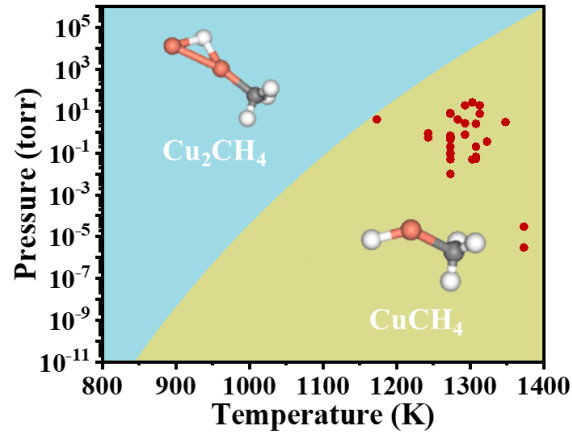

**Fig. 4.** P–T relationship of stable Cu$_x$CH$_y$ (x=1,2, y=0~4) clusters. Red dots represent experimental P–T conditions [11-37].

Further, there are abundant CH₄ and H₂ in the growth chamber under growth conditions. The high surface area and chemical activity of metal clusters enable the formation of complex CuCH ternary clusters, which may act as potential reaction intermediates under experimental conditions. This prompts important questions about the stability and proportions of these clusters, requiring further thermodynamic analysis to link stability with the P–T conditions of growth. Given the critical role of $Cu_1$ and $Cu_2$ in cluster catalysis, probable $Cu_xCH_y$ (x=1,2, y=0~4) cluster structures were carefully considered (see Sec. 3 of the Supplemental Material for more details). The Gibbs free energy of $Cu_xCH_y$ clusters can be expressed by the following formula:

$$G = E_{total} - N_{Cu}\mu_{Cu} - N_C\mu_C - N_H\mu_H \quad (12)$$

For each cluster, its expression of $G(T, P, \chi)$ can be derived by incorporating equations (5), (10), and (11) into (12). Notably, only $CuCH_4$ and $Cu_2CH_4$ clusters appear in the phase diagram as depicted in Fig. 4. The $CuCH_4$ cluster predominantly occupies the phase diagram at higher temperatures and lower pressures. Interestingly, the red dots in the diagram, which represent the P–T conditions observed during experiments, are predominantly located in this region. This suggests that the $CuCH_4$ cluster is likely the dominant ternary structure following the initial dehydrogenation step.

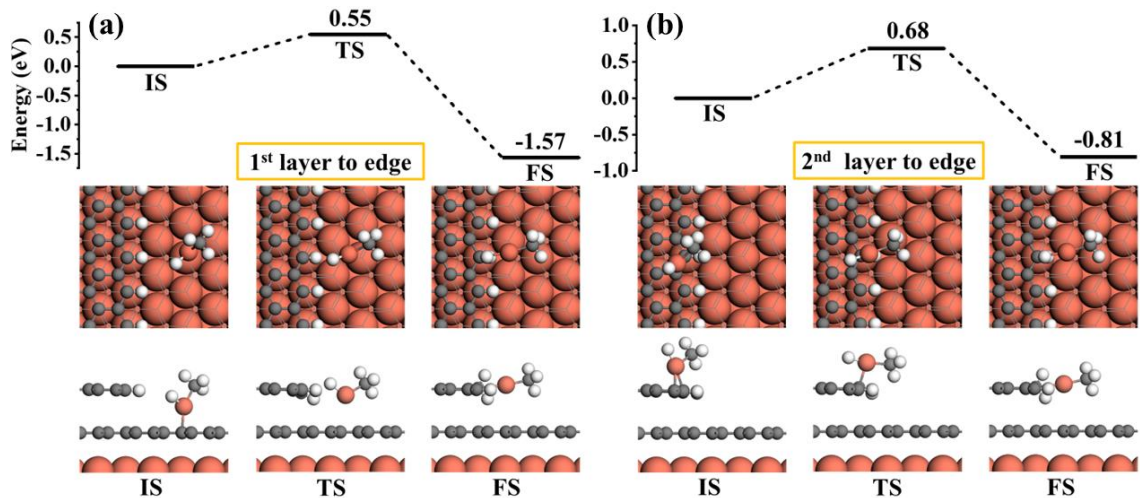

**Fig. 5.** Migration of $CuCH_4$ (**a**) from first layer to edge and (**b**) from second layer to edge.

After recognizing the catalytic reactivity and stability of clusters, exploring their adsorption behaviors and roles in graphene growth became crucial. There is currently clear experimental evidence indicating that graphene on catalytic substrates such as Cu, Ni, and Ru follows a underneath growth mode, where the second graphene layer nucleates and grows at the bottom of the first graphene layer [27, 42]. However, this conclusion does not take into account the influence of the continuous supply of catalytic

Cu vapor, where additional Cu vapor clusters may promote the nucleation at the top of existing graphene. According to our previous works [43-46], nucleation kinetics involves defects, impurities, and steps, making the process highly complex. This complexity warrants further in-depth investigation in future studies.

In this work, we propose a hypothesis that nucleation has already occurred in the second layer and there are stable graphene edges. This is similar to the on-top growth mode, where the second layer of graphene grows on the surface of the first layer of graphene [47-49]. According to our anticipation, the growth of the first graphene layer still relies on the catalytic effect of Cu(111) substrate. Upon completion of the first layer, which shields the Cu surface, the importance of Cu clusters increases. This may lead to layer-by-layer growth of graphene. Thus, a model was constructed featuring graphene supported on Cu(111) substrate, consisting of a complete graphene layer with a localized H-terminated zigzag graphene nanoribbon on the second layer. Subsequent investigations focused on the adsorption of the CuCH$_4$ cluster onto this system:

$$E_{ads} = E_{total} - E_{sub} - E_{cluster} \tag{13}$$

where $E_{total}$ is the total energy of system, $E_{sub}$ and $E_{cluster}$ represent the energy of substrate and cluster, respectively. According to our results as shown in Fig. 5, the adsorption of CuCH$_4$ cluster is strongest at the H-terminated edge. Once the CuCH$_4$ cluster settles on the surfaces of the first and second layers of graphene, it can migrate towards the edge with extremely low barriers of 0.55 eV and 0.68 eV, respectively. The energies of final states are lower than the initial states in both migration paths, indicating the stability of adsorption at the edge. A previous related study pointed out that the passivation of the edge by Cu can reduce the barrier for C atoms to incorporate into the edge from 2.5 eV to 0.8 eV [50]. The above kinetics at the edge of graphene nucleus may represent the upper graphene growing mechanism catalyzed by copper clusters after the Cu substrate is fully covered by graphene. It is important to note that, the underneath growth mode is still widely regarded as the most reasonable explanation for graphene growth. The preceding discussion is predicated on the assumption that stable graphene nuclei reside on the surface of graphene monolayer. Controlled seed introduction may enable the growth of graphene bilayer and multilayer architectures, which requires further experimental evidence.

Our findings elucidated the P–T conditions for Cu vapor formation, its role in $CH_4$ dehydrogenation, and the kinetics of $CuCH_4$ cluster at the edge of graphene nucleus. Even after the Cu substrate is covered with a layer of graphene, the presence of free Cu vapor in the growth chamber may still keep reactivity, probably leading to localized bilayer structures. In previous experimental reports of the growth of bilayer graphene, it was hypothesized that the upstream Cu substrate first dehydrogenates $CH_4$, with methyl radicals flowing downstream to participate in growth [51, 52]. The subsequent dehydrogenation process was attributed to the weak cross layer catalysis of Cu substrate and the complex gas-phase reactions at high concentrations of methyl radicals. There was lack of detailed consideration about the catalytic role of Cu vapor and the correspond surface defects. The study of Cu cluster catalysis deepens the understanding of graphene preparation.

**Methods**

First-principles calculations were performed by using the Vienna ab initio simulation package (VASP) [53]. The generalized gradient approximation (GGA) using the function expressed by Perdew, Burke and Ernzerhof (PBE) was employed to solve the ion-electron interaction [54]. Empirical DFT-D3 (BJ) correction [55] was introduced according to the Van der Waals interaction induced by C and H atoms. Spin polarization effects were considered by turning on the switch of ISPIN=2. The kinetic energy cutoff of 400 eV was test to ensure an accurate convergence for wave functions. The convergence criteria of energy and force were set to $10^{-5}$ eV and 0.01 eV/Å, respectively. The dehydrogenation and H migration processes were calculated by the climbing image nudged elastic band (CI-NEB) method [56]. We inserted 5 images between initial state and final state to find transition state. Except for necessary transition state parameters, all other settings were consistent with structural optimization. All the accuracy of our numerical procedure had been carefully tested.

**Supporting Information**

Additional computational details of Cu cluster formation conditions, CH$_4$ dehydrogenation, and thermodynamic discussions.

**Acknowledgment**

The authors acknowledge the financial support provided by the National Key R&D Program of China (2024YFA1409600) and the National Natural Science Foundation of China (Grant No. 12374253, 12074053, 12004064). J. G. thanks the foreign talents in the project (G2022127004L) and discussion with Prof. Z. Yu. The authors also acknowledge Computers supporting from DUT supercomputing center, Shanghai Supercomputer Center and Sugon Supercomputer Center.

# Copper Vapor Catalyzing Role in the Growth of Graphene

Yuan Chang[a§], Shiji Li[a§], Tianyu He[a], Hongsheng Liu[a], Junfeng Gao[a,b]*

[a] *State Key Laboratory of Structural Analysis for Industrial Equipment, Dalian University of Technology, Dalian, 116024, China*

[b] *Suzhou Laboratory, Suzhou, 215123, China*

*Email: gaojf@dlut.edu.cn

[§] Y. C. and S. L. contributed equally to this work

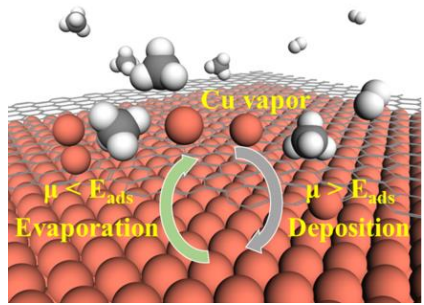

ToC figure.